\documentclass[%
 reprint,
 longbibliography,
superscriptaddress,
 amsmath,amssymb,
 aps,
]{revtex4-1}
\usepackage{graphicx}
\usepackage{dcolumn}
\usepackage{bm}
\usepackage{subfigure}  

\begin{document}

\preprint{APS/123-QED}

\title{Cooling Quantum Gases with Entropy Localization}

\author{F. Nur \"{U}nal}
\email{fatmanur@bilkent.edu.tr}%
\affiliation{Laboratory of Atomic and Solid State Physics, Cornell University, Ithaca, New York 14853, USA}
\affiliation{Department of Physics, Bilkent University, 06800 Ankara, Turkey}
\author{Erich J. Mueller}
\affiliation{Laboratory of Atomic and Solid State Physics, Cornell University, Ithaca, New York 14853, USA}




\date{\today}

\begin{abstract}
We study the dynamics of entropy in a time dependent potential and explore how disorder influences this entropy flow. We show that disorder can trap entropy at the edge of the atomic cloud enabling a novel cooling method. We demonstrate the feasibility of our cooling technique by analyzing the evolution of entropy in a one-dimensional Fermi lattice gas with a time dependent superlattice potential.
\end{abstract}

\pacs{Valid PACS appear here}
\maketitle

\section{Introduction}
Disorder, often treated as a nuisance to be avoided, can be a great resource. For example, the quantum Hall effect is widely believed to only be observable because of disorder \cite{IQHDisorder}. More recently, there have been proposals to use disorder to stabilize topological orders against temperature \cite{DisorderTopology,DisorderTopology2}. Here, we propose a disorder-enabled cooling technique for cold atoms, which takes advantage of the theoretical \cite{LocOrso,LocAbanin,DeMarcoTheory} and experimental \cite{DeMarcoExperiment,BlochMBL} developments involving many-body localization in ultracold atoms.

In discussing ``cooling" of cold atomic systems, the relevant quantity is often entropy rather than temperature \cite{HoSqueezeEntropy,KohlDimple,EntropyScalettar,EntropyShenoy,EntropyWalraven,HoMarch,DeMarcoCooling,HuletCooling, CoolingWithDisorder,MuellerDimple}. Temperature can be radically reduced by adiabatically changing system parameters \cite{HoHeating,IntCooling,FeshbachCooling,AdiabaticCoolFermiLatt} (for example the depth of an optical lattice), but, there is no utility in lowering the temperature if the other energy scales in the system are commensurably reduced. One prevalent idea in the field involves cooling by spatially segregating the entropy \cite{ZwergerUnitaryFermi}. This approach is most thoroughly worked out in the context of dimple traps \cite{KohlDimple}, where a deep potential well yields a low-entropy region in the midst of a shallow trap. Here, we pursue the idea of using disorder to control the spatial distribution of entropy in a trapped atomic cloud.

It is straightforward to create atomic clouds with a central low-entropy region. For example, a Fermi lattice gas with a band insulating core will have most of its entropy at the edge, which is metallic. The low-entropy region, however, is boring. It has a gap to excitations. One needs a way to adiabatically transform the insulating state into something more interesting without allowing the entropy to flow into that region. One set of proposals involves removing the high-entropy atoms while simultaneously changing the confining potential \cite{HoSqueezeEntropy,KohlDimple}. Here, we propose an alternative, namely using disorder to prevent the diffusion of entropy from the edge of the cloud.

Indeed, Anderson showed that, in the absence of interactions, sufficiently strong disorder prevents transport, and would freeze the spatial distribution of entropy \cite{Anderson,Anderson2}. Half a century later, Basko et al. coined the phrase `many-body localization' showing that this insulating behavior survives weak interactions at finite temperature \cite{Basko}. Further experimental and theoretical studies confirmed these results, and showed they persist under very general conditions \cite{LocInguscio,LocBouyer,LocDeMarco,LocAspect, DeMarcoExperiment,DeMarcoTheory,BlochMBL,MattMuller}. One expects that generically disorder can be used to prevent entropy flow, even in the presence of interactions.

To demonstrate our idea, we investigate the dynamics of a simple model of harmonically trapped one-dimensional spin-polarized fermions. A superlattice of period two results in insulating behavior near the middle of the trap and metallic behavior at the edges. Due to the location of the low energy excitations, most of the entropy in the system resides at the edges. We subsequently eliminate the gap in the bulk by ramping down the superlattice potential. This potentially results in a low entropy metallic state for which interactions can lead to novel quantum phenomena. We show that, in the absence of disorder, ramping down the superlattice affects the entropy mainly in two ways. First, due to the harmonic confinement, entropy flows into the center. Second, for finite sweep rates, removing the superlattice potential generates some entropy. We find that sufficiently strong disorder prevents the entropy flow, effectively cooling the central region. We study the entropy dynamics for different sweep rates and compare the degree of entropy localization for different disorder strengths. We also analyze the entanglement entropy in the system to characterize the entropy generation. Finally, we comment on the effect of interactions and experimental considerations.

\section{The Model}   \label{Sec model}
The Hamiltonian of our 1D noninteracting system of spinless fermions can be written as
\begin{multline} \label{hamiltonian}
\frac{\mathcal{H}(t)}{J}=\sum_{i=-N/2}^{N/2} -(a_i^{\dag}a_{i+1}+a_{i+1}^{\dag}a_i)+\frac{1}{2}\omega^2 i^2 a_i^{\dag}a_i \\
+ \Delta(t) (-1)^i a_i^{\dag}a_i + \zeta_i a_i^{\dag}a_i ,
\end{multline}
with nearest-neighbor tunneling rate $J$ and adimensionalized harmonic trap frequency $\omega$. The operator $a_i^{\dag}\,(a_i)$ creates (annihilates) a particle at site $i$. The superlattice strength is parameterized by dimensionless $\Delta$, which we take to be time dependent. For $\Delta\gg 1$, one finds two bands separated by a gap of order $2\Delta$. We introduce uncorrelated disorder $\zeta_i$, uniformly distributed with $|\zeta_i|\leq\zeta$ where $\zeta$ determines the disorder strength. Initially, we assume the system is in thermal equilibrium with chemical potential $\mu$ and temperature $T$. This Hamiltonian can be represented as a matrix. We diagonalize $\mathcal{H}$, finding single-particle eigenstates $\Psi^{(n)}$ and eigenvalues $\varepsilon_n$. The entropy of the system is $S=-\sum_n f_n\ln(f_n)+(1-f_n)\ln(1-f_n)$ where $f_n=(1+e^{(\varepsilon_n-\mu)/kT})^{-1}$ is the Fermi-Dirac distribution. We find it convenient to not include Boltzmann's constant. It is then natural to introduce a local entropy density
\begin{equation}\label{Eq entropy}
S_i=-\sum_{n} |\Psi_i^{(n)}|^2 \left(f_n\ln(f_n)+(1-f_n)\ln(1-f_n)\right),
\end{equation}
so that $S=\sum_i S_i$. As we discuss later, this von Neumann definition does not capture entropy associated with quantum entanglement. For thermal ensembles, however, it is a good definition. In our simulations, we take $N=200$ sites, and tune the gap $\Delta$, trap frequency $\omega$ and chemical potential $\mu$ so that the system supports metallic excitations at the edges with a bulk insulator in between.

\begin{figure}
\centering
\subfigure{\includegraphics[width=0.48\textwidth]{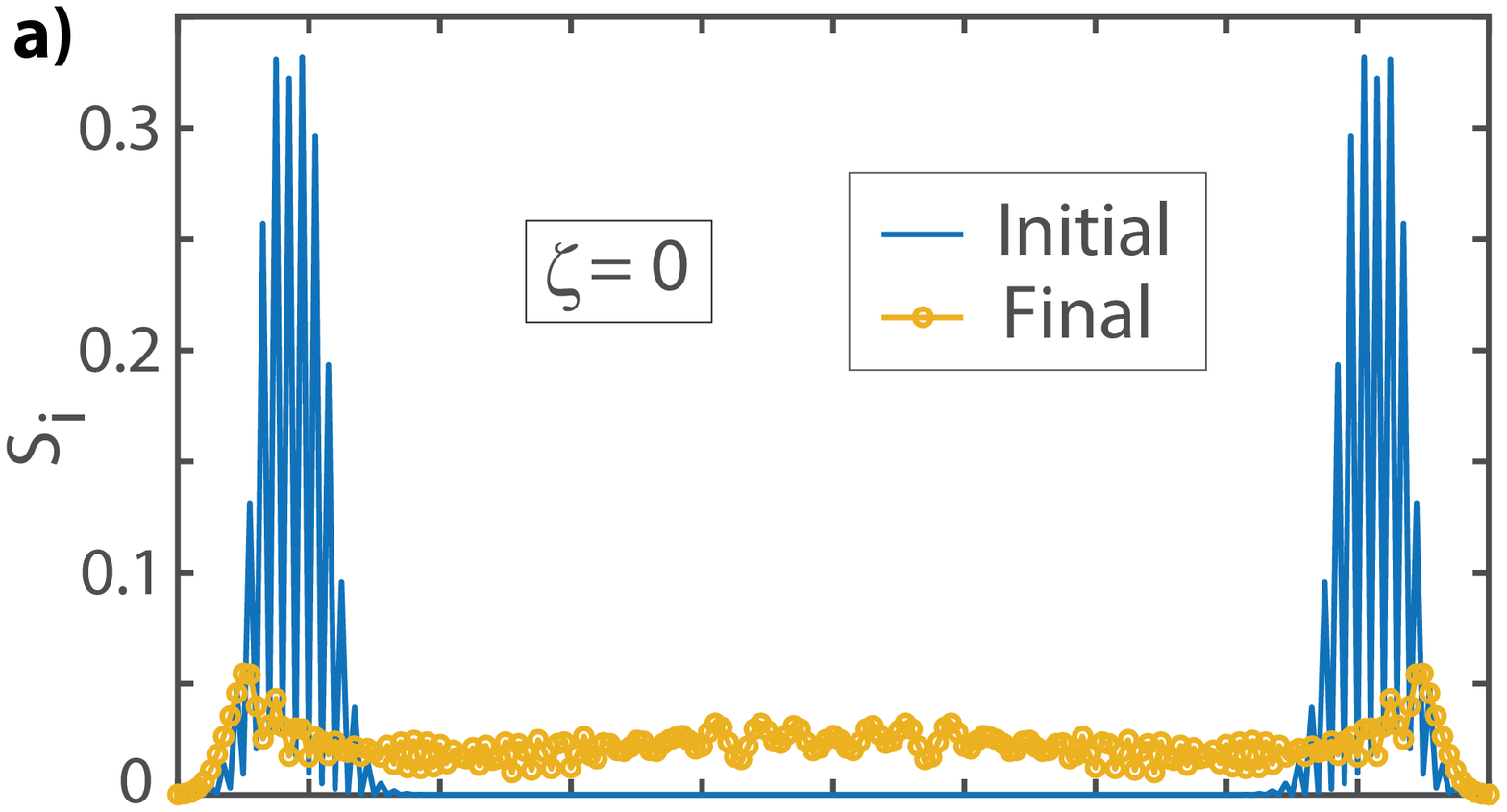}}
\subfigure{\includegraphics[width=0.48\textwidth]{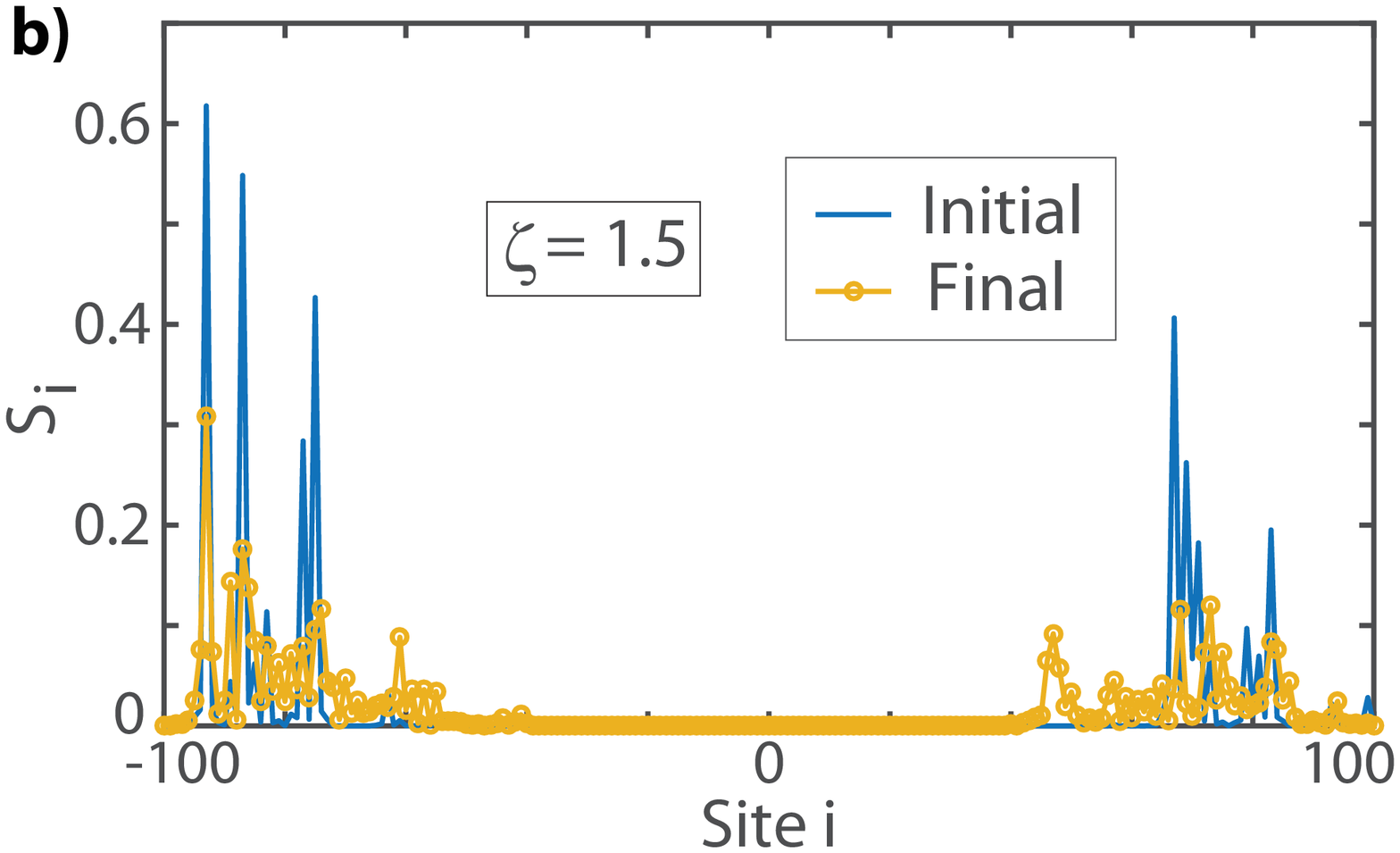}}
\caption{ Local entropy density defined by Eq.(\ref{Eq entropy}) for superlattice strength $\Delta=2.5$, trap frequency $\omega=0.03$, chemical potential $\mu=0.23$ and temperature $T=0.1$. The parameters are given in dimensions of the tunneling rate $J$. The dark lines correspond to initial equilibrium distribution in the presence of a superlattice potential. The light circles show $S_i$ after ramping down of the superlattice potential over a time $\tau=600$. a) In the disorder free case, entropy flows in from the edges as the superlattice potential is turned off. b) Strong disorder prevents this flow by localizing the entropy at the edges. }\label{Sx figs}
\end{figure}

We study how the entropy density evolves with time. In any isolated quantum system (interacting or non-interacting) the total entropy cannot change: A pure state cannot evolve into a mixed state. Regardless of how adiabatic the evolution is, no information is lost in quantum dynamics. Hence, no unitary evolution can change the von Neumann entropy in an isolated system. The spatial distribution of the entropy, can however evolve. We will largely be considering a non-interacting gas, where the occupation factors $f_n$ in Eq. (\ref{Eq entropy}) will be constant, but the wave functions $\Psi_i^{(n)}$ may evolve with time. This time-dependent Hartree-Fock approximation, which was first proposed by Dirac \cite{dirac1930}, is exact for a non-interacting gas. However, even in the case of interactions, it is accurate for describing modes which have frequencies large compared to the inverse collision time.

Physically we expect that, given enough degrees of freedom, an isolated quantum system should be capable of thermalizing \cite{ThermalGoldstein,ThermalPolkovnikov,ThermalTasaki,ThermalDeutsch,ThermalSred1,ThermalSred2}. Thermalization requires entropy growth, so this physical expectation is at odds with the mathematical statement that the entropy is constant. One solution to this puzzle is to consider the {\em entanglement entropy} of a subregion (see Section \ref{Sec entglmt entropy} and Ref. \cite{DeutschNJP}). For generic quantum states the entanglement entropy of a small subregion is proportional to the volume of that region, allowing one to define a quantum entropy density. This quantum entropy density generically increases with time. The total entropy, as conventionally defined, is not equal to the volume integral of this quantum entropy density. There are alternative procedures which allow one to define entropy densities which increase with time in isolated systems \cite{Ueda,EntQuench,Polkovnikov}.

In Section \ref{Sec entropy density}, we explore the entropy redistribution, as captured by Eq. (2). In section \ref{Sec entglmt entropy}, we calculate the evolution of the entanglement entropy of the central region. These are both valid ways of defining entropy density, and reveal different aspects of the dynamics. We show that regardless of the definition of entropy, the disorder reduces the entropy growth in the center of the cloud.

\section{Results} \label{Sec Results}
\subsection{Entropy Density}  \label{Sec entropy density}
The dark blue lines in Fig. \ref{Sx figs} show the initial entropy density with and without disorder. Clearly, the entropy is initially concentrated at the metallic edges. One hopes that the low entropy density at the center of the trap can be used as a resource. As previously explained, in order to make use of this resource we need to eliminate the gap by reducing $\Delta$ to zero. Thus, we wish to calculate how the entropy evolves as we change the superlattice strength. In the absence of scattering, we can use the single-particle Schr\"{o}dinger equation to evolve the wave functions, keeping the occupation factors fixed. We assume a linear ramp,
\begin{equation}
\Delta(t)=
\begin{cases}
\Delta_0-\frac{\Delta_0}{\tau}t, & \quad 0\leq t\leq\tau,\\
0, & \quad t>\tau.
\end{cases}
\end{equation}
where larger $\tau$ corresponds to a slower sweep. In the disorder-free case, entropy defined by Eq. (\ref{Eq entropy}) flows in from the edges as we close the gap. This behavior is reasonable as we know a fully adiabatic ramp would result in a thermal state, whose entropy density is peaked at the center of the cloud. We caution, however, that true adiabaticity requires extremely slow sweeps. The flow of entropy towards the center is nonetheless robust, occurring even in relatively fast sweeps. Fig. \ref{Sx figs} shows that, as anticipated, strong disorder ($\zeta=1.5$) localizes the entropy at the edge of the cloud during the evolution. Although the local entropy density is low, the state is nominally non-thermal. The states $\Psi^{(n)}$ at the final time are not energy eigenstates. Nonetheless, in the central region, the system will behave in many ways similar to a low temperature state. The fluctuations will be small.

\begin{figure}
\includegraphics[width=0.48\textwidth]{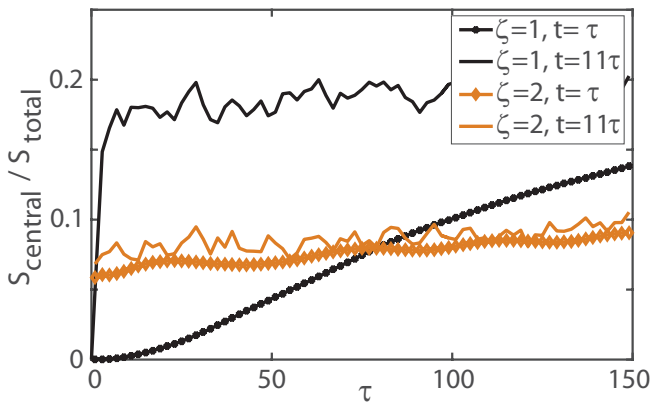}
\caption{ The fraction of the entropy in the central region of the trap ($-60<i<60$ for $N=200$ sites). Here, the superlattice strength is $\Delta=3$, trap frequency is $\omega=0.035$, chemical potential is $\mu=0.75$ and temperature is $T=0.1$. The parameters are given in dimensions of the tunneling rate $J$. The dots and the diamonds correspond to entropy immediately after the sweep $t=\tau$ and the solid lines correspond to $t=11\tau$ where we allow the system to evolve further after the sweep is complete. We show two different disorder strengths, $\zeta=1$ (dark) and $\zeta=2$ (light). For weaker disorder, there is significant entropy flow following an abrupt ramp, so to achieve the adiabaticity the ramp must be slower.} \label{S Central fig}
\end{figure}

We find that the entropy evolution is sensitive to sweep rate ($1/\tau$). In a fast sweep (small $\tau$) where the wave functions do not have enough time to adjust themselves to the new Hamiltonian, the entropy distribution immediately after the sweep would be similar to the initial configuration, i.e. trapped at the edges. Fig.\ref{S Central fig} demonstrates these dynamics at time $t=\tau$ for two different disorder strengths, $\zeta=1$ (dots) and $\zeta=2$ (diamonds), and the entropy is initially concentrated at the edges. We consider the relative percentage of the entropy that resides in the center of the trap (i.e. between $-60<i<60$ for $N=200$ sites). This central region holds $75\%$ of the particles. Strong disorder ($\zeta=2$) enhances the adiabaticity of the process and the central entropy percentage becomes largely independent of sweep rate. However, for weaker randomness ($\zeta=1$), the central entropy seems to increase initially as we make the sweep slower and then saturates to a finite value.

\begin{figure}
\includegraphics[width=0.48\textwidth]{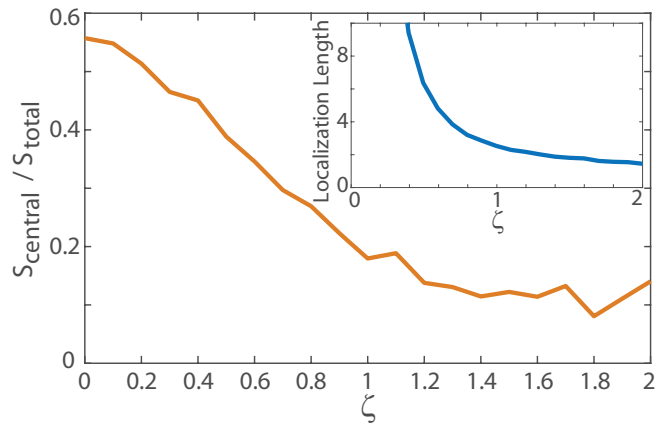}
\caption{ The fraction of the final entropy in the central region of the trap ($-60<i<60$ for $N=200$ sites) vs. disorder strength. The superlattice strength is $\Delta=3$, trap frequency is $\omega=0.035$, temperature is $T=0.1$, and chemical potential is fixed at $\mu=0.75$. We take $\tau=100$ and let the system evolve for another $10\tau$ after ramping down the superlattice. Initially for a clean system, $56\%$ of the total entropy lies in the central region. Increasing disorder quickly freezes the entropy at the edges. The inset displays the corresponding localization lengths. When the localization length is around 2 lattice sites, the central entropy percentage is already reduced to a third of the disorder-free case. }
\label{Scentral vs Dis. fig}
\end{figure}

One important concern is that the system continues to evolve following the sweep with entropy continuing to spread towards the center. In order to study this effect, we let the system evolve for another $10\tau$ after the sweep is completed, i.e. the total time of the evolution is $11\tau$. For weaker disorder strength, the entropy evolves significantly after the sweep. After a long time, the central entropy density is nearly independent of sweep rate, saturating near $18\%$ for $\zeta=1$. A considerable percentage of the entropy still remains frozen at the edges of the cloud.

For strong disorder, the entropy, as defined by Eq. (\ref{Eq entropy}), fails to evolve following the sweep. Moreover, the amount of entropy which flows in during the removal of the superlattice potential decreases as the disorder increases. For $\zeta=2$, only $10\%$ of the total entropy flows into the middle of the trap. We consider this dependence of the final central entropy on the disorder strength in Fig. \ref{Scentral vs Dis. fig}. In order to analyze the strength of the disorder, we also display the corresponding localization length in the inset of Fig. \ref{Scentral vs Dis. fig}, which is calculated by analyzing the exponential tails of the wave functions \cite{AspectLocLength,LocAspect}. In the disorder-free case, almost $60\%$ of the total entropy resides in the center following the sweep, which is compatible with the length of this region. When the localization length is around two lattice sites, the central entropy percentage is already reduced to a third of the disorder-free case. In fact, for the parameters given in Fig.\ref{S Central fig} and Fig.\ref{Scentral vs Dis. fig}, the entropy per particle is reduced by a factor of 3 to 10 in the center. These results prove that when the system is pre-cooled with conventional techniques, our disorder-induced cooling mechanism can be employed to reach temperatures much lower in the center than the rest of cloud. This is particularly useful in obtaining low temperatures in optical lattice systems or in the presence of speckle disorder.

The fine structure noise in Fig.\ref{S Central fig} and Fig.\ref{Scentral vs Dis. fig} has two sources. First, there are rapid oscillations associated with particular disorder realizations. We somewhat control these by averaging over thirty realizations. Second, there are longer wavelength wiggles in Fig.\ref{S Central fig} which are associated with the trap.

\subsection{Entanglement Entropy}  \label{Sec entglmt entropy}
The definition of entropy in Eq.(\ref{Eq entropy}) does not capture any entropy generation during the ramping down of the superlattice potential. One convenient way to characterize any entropy generation is to look at the entanglement entropy between the central region and the rest of the cloud \cite{Sent1,Sent2}. For our state, this entanglement entropy can be calculated from the single particle density matrix,
\begin{equation}
G_{ij}=<\hat{\Psi}_i^{\dag}\hat{\Psi}_j>=\sum_n \Psi_i^{(n)*}\Psi_j^{(n)}f_n,
\end{equation}
where $i$ and $j$ label sites. Cheong and Henley  showed that if one truncates this matrix, restricting $i$ and $j$ to lie in a subregion, then the entanglement entropy is related to the eigenvalues $\lambda_m$ of the truncated density matrix \cite{ChrisHenley}. In particular,
\begin{equation} \label{Eq S entanglement}
S_{entanglement}=-\sum_m \lambda_m \ln(\lambda_m)+(1-\lambda_m)\ln(1-\lambda_m).
\end{equation}

\begin{figure}
\includegraphics[width=0.48\textwidth]{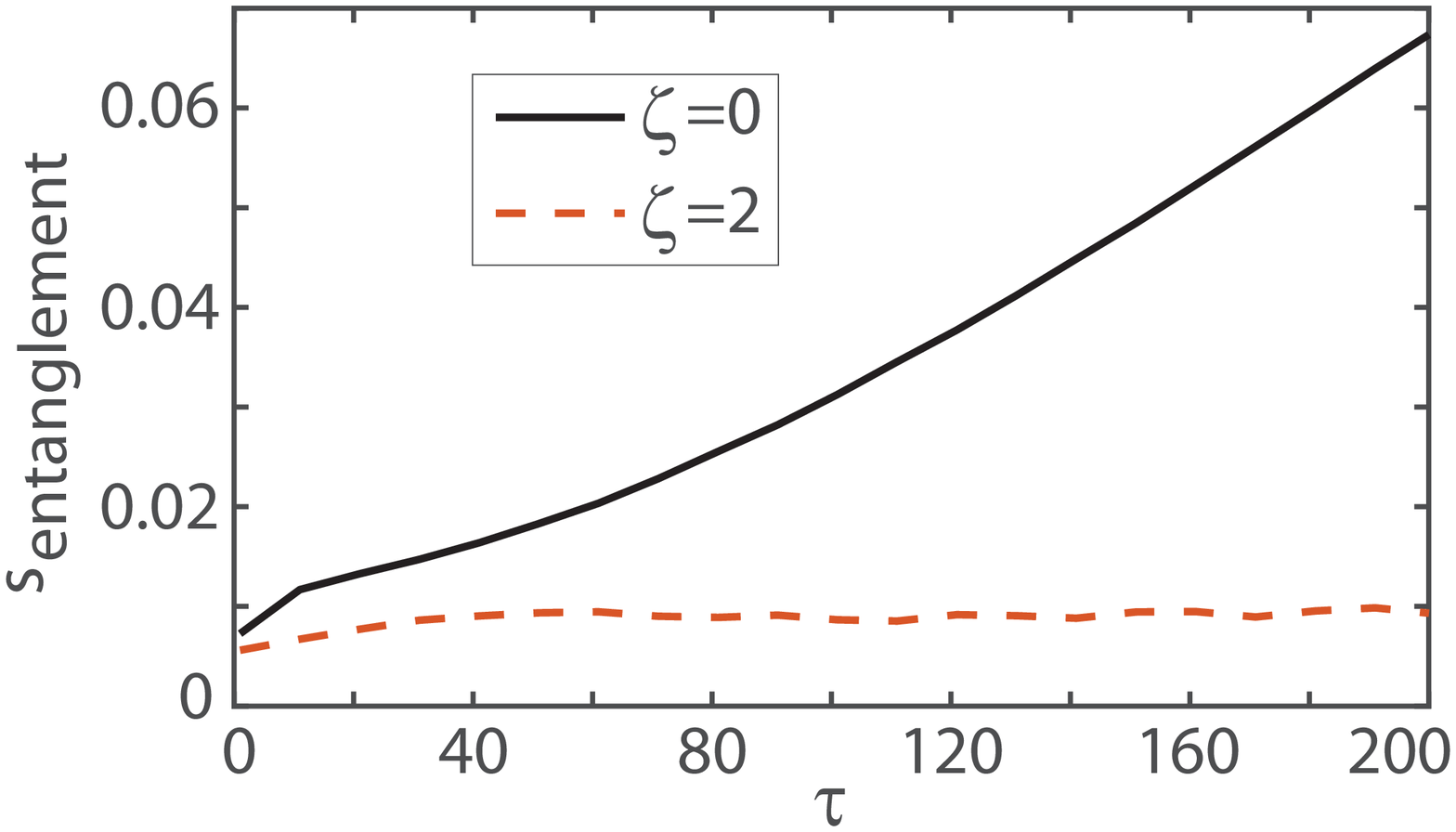}
\caption{ Dimensionless entanglement entropy per site between the central region and the rest of the cloud for superlattice strength $\Delta=3$, trap frequency $\omega=0.035$ and temperature $T=0.1$ immediately following a sweep of duration $\tau$. The parameters are given in dimensions of the tunneling rate $J$. The entanglement entropy becomes small as $\tau$ goes to zero because there is less time for information to propagate. In principle, the entropy should again be small at very large $\tau$ when the dynamics are truly adiabatic. In the disordered case (dashed line), localization limits the amount of entanglement possible.  }
\label{S_ent fig}
\end{figure}

$S_{entanglement}$ measures how much the central region becomes correlated to the rest of the system while the superlattice is being ramped down. For our calculation, we consider the entanglement entropy of the center of the cloud, taking $-60<i,j<60$ for $N=200$ sites. In Fig.\ref{S_ent fig}, we demonstrate the central entanglement entropy per site ($s_{entanglement}=S_{entanglement}/120$) for the disorder free case and the strong disorder. Initially, the central entanglement entropy density is almost zero (not displayed in Fig.\ref{S_ent fig}) for both cases. In the absence of disorder, $s_{entanglement}$ immediately after the sweep increases for increasing $\tau$ and then saturates to a finite value. This increase again reflects continuing evolution of the entropy after an abrupt ramp. In principle, for infinitely slow sweeps no entropy will be generated. For practical sweep rates however, we find that more entropy is generated for slower sweeps. This is in part because longer sweeps provide more time for the entropy to evolve. As one expects, adding disorder suppresses entropy generation for slower sweeps. Fig.\ref{Sx figs}-\ref{S_ent fig} demonstrate that both the entropy flow and the entropy generation can be suppressed by using disorder.

\section{Conclusion and Outlook}
Cooling atomic gases down to temperatures low enough to observe novel quantum phenomena is an ever present challenge. The current cooling techniques mostly rely on removing the high entropy particles from the system \cite{StamperKurnCooling}, which usually lie at the edges of the system. Instead, we propose a cooling technique where disorder is used to control the spatial distribution of entropy. In particular, we demonstrate our disorder-induced cooling mechanism by applying it to one-dimensional non-interacting fermions in a harmonic trap. By employing a period two superlattice, we create a gap in the spectrum and a low entropy region in the center of the cloud. Introducing disorder to the system localizes the entropy at the edges. We then adiabatically remove the superlattice potential to obtain a metallic low-entropy state at the center and analyze the dynamics of the entropy during the evolution. We show that only a small percentage of the total entropy lies in the central region. Since the system has been already cooled down with conventional means before ramping down the spectral gap, the central low-entropy region can then provide access to temperatures much lower than the rest of the cloud \cite{KohlDimple}.

Our ideas are particularly valuable for producing very cold disordered gases. Typically it is extremely hard to cool in lattices or speckle disorder \cite{DeMarcoCooling}. Our approach, where a superlattice potential is ramped down in the presence of disorder overcomes these difficulties, providing a promising way to create a disordered low entropy gas.

Although we model the case of a superlattice potential here, our approach should work in much general settings. The only requirement is that there is a spectral gap in the center of the cloud, with gapless excitations on the edge. One adds disorder to the system and cools as far as possible with conventional means. One then slowly changes the Hamiltonian to turn off the central gap. One could also imagine interesting variants, where the disorder is only applied to the edge of the cloud so that one would have a homogenous system in the center.

Our disorder-induced cooling mechanism can be combined with existing cooling techniques to further lower the temperatures in these systems. For example, after using disorder to trap the entropy at the edges, one can use the techniques from Ref.\cite{HoSqueezeEntropy,KohlDimple} to remove these high-entropy particles from the system. Once the atoms at the edges are separated from the center, one can think about other modifications depending on the particular system at hand. For example, Ref.\cite{CoolingWithDisorder} introduced another cooling technique by adiabatically ramping down the disorder with the aim of reaching the N\'{e}el temperature, however, the technique was not sufficient on its own and required an additional scheme to reduce the entropy initially. Our cooling mechanism is a promising candidate for this pre-cooling. For the parameters given in Fig.\ref{S Central fig}, we find roughly a factor 10 reduction in temperature, which can be sufficient to reach the N\'{e}el transition. However, more work is needed to understand the interaction between the motional degrees of freedom studied here, and spin. Ramping down the disorder is also appealing in that it provides a clean homogeneous system.

Our calculations neglect interparticle interactions. We expect, however, that our results are robust. Interactions profoundly change the behavior of the clean system: collisionless ballistic motion is replaced by diffusion. In the disordered system, however, the role of the interactions are much more subtle. Extensive theoretical work shows that even when pushed far from equilibrium, the disordered interacting system displays localization \cite{HuseReview,MooreLocArxiv}. Thus, even in the presence of interactions, we expect disorder will trap entropy at the edge of the cloud. Modeling the dynamics of the interacting system is much more involved, and will be reserved for future studies.

\section{Acknowledgements}
F.N.\"{U}. is supported by The Scientific and Technological Research Council of Turkey (T\"{U}B\.{I}TAK). This material is based upon work supported by the National Science Foundation under Grant No. PHY1508300.

\bibliography{EntropyNJP}

\end{document}